\documentclass{agnspec}
\usepackage{graphics}
\usepackage{psfig}
\def\ion2#1#2{#1$\,${\small\rm{#2}}\relax}
\begin{document}
\title{A {\it Chandra} Observation of the Luminous NLS1 1H 0707$-$495}
\author{Karen M.\ Leighly\inst{1}, Andrzej A.\ Zdziarski\inst{2}, Toshihiro
Kawaguchi\inst{1,3}, and Chiho Matsumoto\inst{1}}
\institute{Department of Physics and Astronomy, The University of
Oklahoma, 440 W.\ Brooks St., Norman, OK 73019
\and N. Copernicus Astronomical Center, Bartycka 18, 00-716 Warszawa,
Poland
\and LUTH, Observatoire de Paris, Section de Meudon, 5, Place Jules Janssen, 92195 Meudon, France }
\authorrunning{K.~M.~Leighly~et.\ al.}
\titlerunning{A {\it Chandra} Observation of the Luminous NLS1 1H 0707$-$495}
\maketitle

\begin{abstract}

We present preliminary results from a long {\it Chandra} HETG
observation of the luminous Narrow-line Seyfert 1 galaxy
1H~0707$-$495.  We find a complex X-ray spectrum comprised
of a two-component continuum with superimposed emission and absorption
lines.  The short time scale X-ray variability is different than
observed in other AGN: the soft X-rays vary markedly less than the
hard X-rays. This behavior is similar to that of high-state Galactic
black holes.  We also investigated the long time scale variability,
and discovered an apparent bimodal flux distribution.  We postulate
that the bimodality is the signature of the radiation pressure
instability, and note that this instability may be expected in
luminous NLS1s.

\end{abstract}

\section{Introduction}

Narrow-line Seyfert 1 galaxies (NLS1s) are subclass of active galaxies
(AGN) identified by their optical emission-line properties.  They have
narrow permitted lines (H$\beta$ FWHM $<2000\, \rm km\,
s^{-1}$), weak
forbidden lines (\ion2{O}{III}/H$\beta<3$), and they frequently show
strong optical \ion2{Fe}{II}.

New attention was focused on this class of AGN when it was discovered
that their X-ray spectrum, both in the soft X-rays observed by {\it
ROSAT} (Boller, Brandt \& Fink 1996), and in the hard X-rays observed
by {\it ASCA} (e.g.\ Leighly 1999b and references therein), is
significantly steeper than that of Seyfert 1 galaxies with broad
optical lines.  It was also discovered that NLS1s show more rapid X-ray
variability than broad-line objects, and sometimes large amplitude
flares are seen (e.g.\ Leighly 1999a).

The origin of these characteristic properties is not yet known.  Early
on, it discovered that the NLS1 RE 1034+39 had an X-ray spectrum
composed of a strong soft excess and a weak steep power law that
strongly resembled that of Galactic black hole objects in their high
state.  From this, Pounds, Done, \& Osborne (1996) proposed that NLS1s
are high-state active galaxies.  Since then, it was recognized that the
X-ray properties are correlated with the optical/UV properties (the
so-called Eigenvector 1; e.g. Boroson \& Green 1992), and all may be
driven by a high accretion rate relative to Eddington.

While this scenario describes the general properties of the X-ray
spectrum and variability well in a schematic way, there are also a
number of uncertainties:

\begin{itemize}
\item The strong soft excess in Galactic black hole candidates can be
directly interpreted as mildly Comptonized blackbody emission from an
accretion disk around a non-rotating black hole.  This simple model is
problematic for NLS1s, as the temperature expected from the standard
disk model and large black hole mass is far too low to be seen in
the X-ray band.
\item Many Galactic black holes in the soft state have variable hard
X-rays and reduced soft X-ray variability (e.g.\ Churazov, Gilfanov,
\& Revnivtsev 2001).  But NLS1s most often show little color
variability, and most frequently their spectra become a little softer
when brighter (e.g.\ Leighly 1999a).
\item The X-ray properties of NLS1s do not seem to be uniform as a
class (Leighly 1999b).  While most objects show a soft excess, the
fraction of energy in that component compared with the power law
spans a wide range, and soft excess strength is correlated with
variability amplitude.  Furthermore, several of the objects with the
strongest soft excess have a peculiar absorption feature near 1 keV
(Leighly et al.\ 1997).
\item NLS1s are common in soft X-ray selected samples (e.g.\ Grupe et
al.\ 1999).  This implies that they are strong soft X-ray sources, and
soft X-rays dominate the spectrum. However, among
NLS1s there are a handful of higher luminosity objects that appear to
be remarkably weak in soft X-rays.  An example is PHL 1811, a very
luminous NLS1 that is anomalously weak in X-rays (Leighly et al.\
2001).  
\end{itemize}

\begin{figure*}
\centerline{\psfig{figure=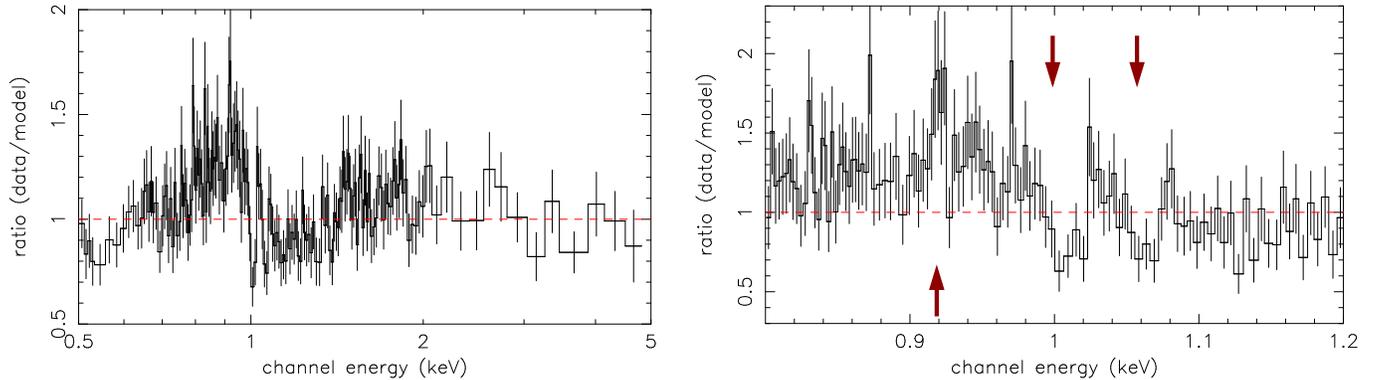,width=18cm}}
\caption[]{The {\it Chandra} MEG first order spectrum from 1H
0707$-$495 (observed energies), binned to 25 photons in each channel.
{\it Left:} The ratio of the data to power law plus black body model.
{\it Right:} Zoomed view, showing significant spectral features.}
\end{figure*}

1H~0707$-$495 ($\rm z=0.041$) is a particularly interesting NLS1.
{\it ASCA} observations of this object revealed a strong soft excess,
high amplitude variability, and an apparent absorption feature around
1 keV (Leighly et al.\ 1997; Leighly 1999ab).  We observed
1H~0707$-$495 with {\it Chandra} in order to identify the 1 keV
absorption feature.  In this contribution, we describe the results,
preliminary analysis, and tentative interpretations of the
observation.

\section{The {\it Chandra} Observation}

The proposed 100~ks observations of 1H~0707$-$495 was made in two
segments, 34~ks and 55~ks in length, that were separated by 15 hours.
1H~0707$-$495 was bright during the observation.  The
average flux was approximately 1.6 times that of the 1995 {\it ASCA}
observation, and the shape of the spectrum seems very similar as
well.  We obtained 9025 photons in the MEG first order spectra between
0.5 and 5 keV.

We obtained about the same number of photons in the zeroth order
between 0.5 and 10 keV, yielding a count rate of 0.1 count/s.  We
estimate that the zeroth order spectrum is about 10\% piled up,
implying that it can be analyzed.  We have commenced analyzing the
spectrum using the {\sc ISIS} package, and the analysis is still
underway.

\begin{figure}
\centerline{\psfig{figure=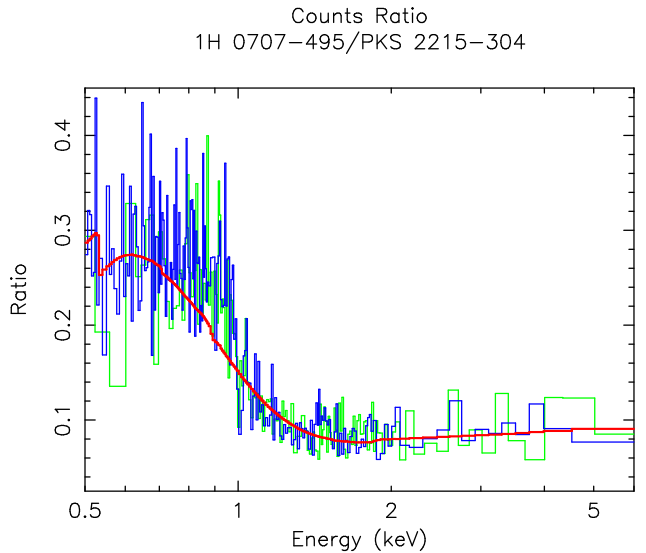,width=8.4cm}}
\caption[]{The ratio of the MEG+1/$-$1 spectra from 1H~0707$-$495 to
those from the calibration source PKS~2155$-$304. Overlaid is the
ratio of the best-fitting continuum models.  This figure demonstrates
that the structure in the 0.8--1.4 keV band cannot be attributed to
residual calibration uncertainties.}
\end{figure}

\section{The {\it Chandra} Spectrum}

The summed MEG first order spectrum was first fitted with a power law,
black body, and Galactic absorption.  The fit was not acceptable,
yielding $\chi^2=426$ for 339 degrees of freedom (d.o.f.).  
The cause of the poor fit is z-shaped continuum residuals between
0.8 and 2 keV (Fig.\ 1).  These occur because the spectrum suffers a
sharp drop at around 1 keV, and this drop is too steep to be modelled
by a black body.

Several discrete emission and absorption line-like features also
contribute to the poor fit.  Given the poor signal to noise, many of
the features are detected with low confidence.  An exception is the
absorption line appearing at 1.046 keV in the rest frame.  This line
has an equivalent width of 6 eV (depending on how the continuum is
modelled) and appears to be marginally resolved.

The spectral modelling is not yet complete.  However, we find
that the following probably contribute:
\begin{itemize} 
\item The effective areas of the different types of CCDs in the HETG
grating spectrometer do not agree (H.\ Marshall 2001, p.\ comm.)  In
particular, a drop in the spectrum between 0.9 and 1.4 keV may be
caused by low effective area for the back-illuminated CCDs relative to
the front-illuminated ones.  To divide out this calibration
uncertainty, we plot the flux ratio of the 1H~0707$-$495 spectrum with
that of the calibration source, the BL Lac PKS 2155$-$304 (Fig.\ 2).
The ratio of the best fitting models are also shown ({\sc pl+bb} for
1H~0707$-$495, and {\sc pl} for PKS 2155$-$304).  This plot shows that
the sharp decrease near 1 keV remains, and is not simply a consequence
of calibration uncertainties.  Adjusting the ancillary response file
(ARF) to account for the calibration differences (H.\ Marshall 2002,
p.\ comm.) reduced the $\chi^2$ by 42 (339 dof).
\item We used a power law plus black body to model the continuum.  In
reality, if the power law is Comptonized black body photons, it should
break toward low energies. This would result in a sharper break at the
junction between the soft excess and power law.  We modelled the
spectrum using the EQPAIR Comptonization model (e.g.\ Gierli\'nski et
al.\ 1999) and found
a reduction in $\chi^2$ by 15.
\item Emission and absorption features contribute to the spectrum
around 1 keV, as suggested by the line-like features shown in Fig.\ 1.
Why there no strong absorption features at lower energies?  Nicastro,
Fiore \& Matt (1999) proposed that the steep X-ray spectrum will
overionized light elements with opacity at low energies, leaving only
highly-ionized iron ions that have significant opacity above 1 keV.
Furthermore, iron L can contribute line emission below 1 keV. The
combination of emission and absorption features may contribute to the
sharp drop.  We are in the process of modelling this situation using
XSTAR, and preliminary results indicate significant improvement when
emission and absorption are considered.
\end{itemize}

\begin{figure}
\centerline{\psfig{figure=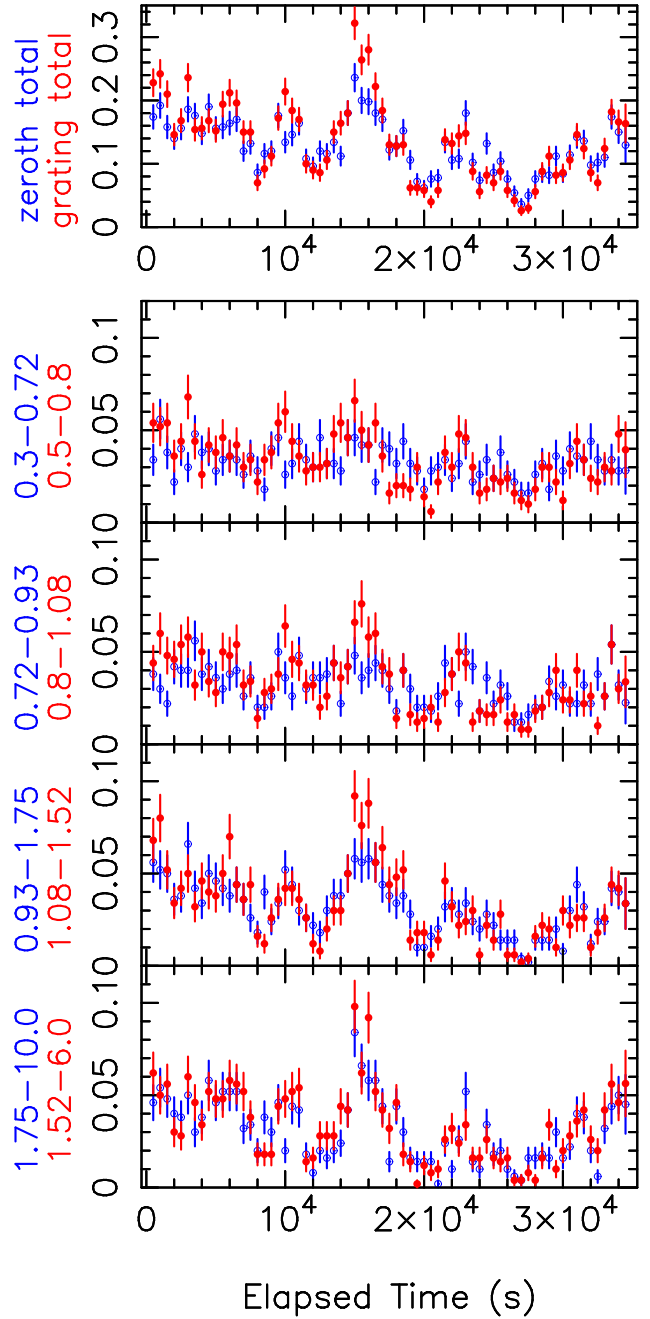,width=8.4cm}}
\caption[]{Light curves from the first observation.  Red filled points
are from the MEG first order grating data, and blue open points are
obtained from the zeroth order data.  The top panel shows the total
flux, and the bottom four panels show the light curves split into
energy bands (marked on the y-axis) such that the total number of
photons is the same in each one. A similar result was obtained for the
second observation.}
\end{figure}

In November 2000, 1H~0707$-$495 was observed using {\it XMM-Newton}
(Boller et al.\ 2002).  During this observation, a spectacular sharp
absorption feature was observed at 7.1 keV in the rest frame.  When
modelled simply as an edge, Boller et al.\ derive an optical depth of
$\tau=1.8 \pm 0.3$. We cannot observe this region of the spectrum
using the gratings; the sensitivity for this low-flux object is much
too low.  However, we can observe it using zeroth order.  We find that
an edge is not statistically necessary and $\tau < 0.8$ (90\%
confidence for one parameter of interest).  We caution that we have
not yet taken pileup into account.  As noted above, the count rate is
sufficiently low that pileup is not expected to present a severe
problem; however, it could artificially decrease the depth of an edge.

\section{Short-term Variability}

During the 40 ks observation {\it XMM-Newton} observation, large
amplitude variability was observed (Boller et al.\ 2002).  The
fractional amplitude of variability was 40\%, and the variability was
color-independent.  During the {\it Chandra} observation, a strikingly
different result was obtained: the soft X-rays are much less variable
than the hard X-rays.  This is clearly seen in the light curves (Fig.\
3), and can also be seen by the decrease in mean-normalized variance
toward lower energies (Fig.\ 4).

As discussed in Section 3, the X-ray continuum from 1H~0707$-$495 can
be grossly fit with two components: a power law and a soft excess. It
is possible that reduced variability in soft X-rays is a consequence
of a variable power law normalization and a stationary soft excess
(red curve in Fig.\ 4).  The figure shows that this simple hypothesis
is ruled out, because it predicts too little variability at low
energies.  It is possible that the variable hard X-ray continuum has
curvature, steepening toward soft X-rays.  Physical origins could be
reflection from an ionized disk, or intrinsic curvature.
Alternatively, the component emitting the soft X-rays may itself vary
at a lower amplitude.  Detailed time series analysis, now underway,
may allow us to distinguish between these possibilities.

\begin{figure}
\centerline{\psfig{figure=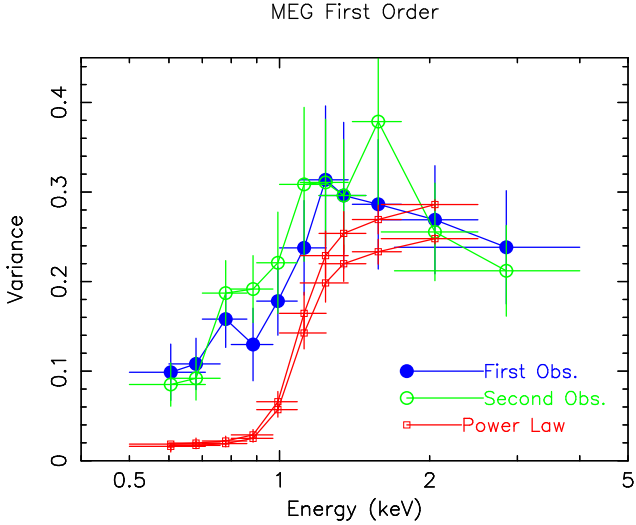,width=8.4cm}}
\caption[]{The mean-normalized variance as a function of energy from
the MEG first order data, for the first and second observations. To
show detail, overlapping energy bands were used; thus, adjacent points
are not independent.  The red curve shows the predicted variance as a
function of energy assuming that the only the power law component
varies.  Two different normalizations for the power law variance are
shown.}
\end{figure}

\begin{figure}
\centerline{\psfig{figure=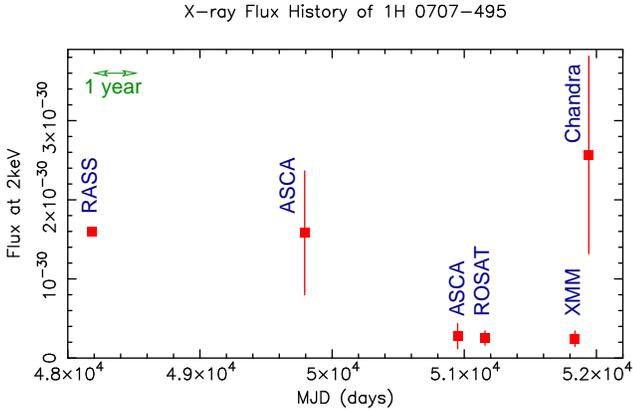,width=8.4cm}}
\caption[]{The flux at 2 keV in the rest frame for 1H~0707$-$495.  The
error bars mark the fractional amplitude of variability rather than
uncertainty; this value is not known for the ROSAT All Sky Survey
point. The {\it XMM-Newton} value was obtained from the plot of the spectrum
in Boller et al.\ 2001.}
\end{figure}

\section{Long-term Variability}

When 1H~0707$-$495 was observed using {\it Chandra}, the average flux
was about 10 times that when it was observed with {\it XMM-Newton}.
Fig.\ 5 shows the long term flux history of 1H~0707$-$495.  The light
curve is quite striking.  Generally, when multiple observations are
available, AGN X-ray fluxes range rather continuously between highest
and lowest states (e.g.\ Grandi et al.\ 1992) and the full range is
generally a factor of a few.  In contrast, although we only have
information for 6 separate epochs for 1H~0707$-$495, we see that the
variability appears to be bimodal, with the difference in high and low
states being about a factor of 10.

We postulate that the bimodal long time-scale X-ray variability is a
consequence of the radiation pressure instability (e.g.\ Honma,
Matsumoto \& Kato 1991).  Why should we see this instability in
luminous NLS1s?  NLS1s are thought to have a high accretion rate,
which implies that a larger portion of the accretion disk is
radiation-pressure dominated.  Furthermore, luminous NLS1s appear to
be generally weak hard X-ray sources, and although part of this should
be due to a cooler corona, the corona may be weak as well.  Some
models predict a weak corona when the accretion rate is high (Liu et
al.\ 2002).  Dissipation of the accretion energy in the corona
effectively stabilizes the disk (e.g.\ Svensson \& Zdziarski 1994), so
a weak corona may also enhance the radiation pressure instability.

The time scale of the instability should depend on black hole mass,
accretion rate, and the viscosity. While different models predict
somewhat different time scales, we find that they may be roughly
consistent the time scales inferred from Fig.\ 5.  For example, the
radiation pressure instability model has been successfully applied to
the variability of the microquasar GRS~1915$+$105 by Janiuk, Czerny,
\& Siemiginowska (2000).  Scaling from the observed time scale of
$\sim 1000$ seconds for a 10$M_\odot$ black hole, we predict a cycle
of 3 years for a $10^6\,M_\odot$ black hole.

\begin{acknowledgements}
We acknowledge support for {\it Chandra} analysis
(GO0--1162X), and NLS1 studies (NAG5-10171).
\end{acknowledgements}

\end{document}